\documentclass[namedreferences,hyperref,optionalrh]{spr-sola}
\usepackage{graphicx}        
\usepackage{color}           
\usepackage{ragged2e} 

\usepackage{indentfirst}

\chardef\us=`\_
\usepackage{amsfonts}
\usepackage{array}
\usepackage{tabularx}
\usepackage{booktabs}
\usepackage{amsmath}
\usepackage{bm}
\usepackage{gensymb}
\usepackage{lineno}
\usepackage{threeparttable}

\usepackage{ragged2e}
\definecolor{DarkGreen}{rgb}{0.0,0.4,0.0}  

\usepackage[normalem]{ulem}
\usepackage{soul}
\usepackage{cancel}

\begin{document}

\begin{frontmatter}
\title{SARD: A YOLOv8-Based System for Solar Active Region Detection with SDO/HMI Magnetograms}

\author[addressref={aff1},email={jhpan@mail.ustc.edu.cn}]{\inits{J.}\fnm{Jinhui}~\snm{Pan}\orcid{}}

\author[addressref={aff1,aff2},email={ljj128@ustc.edu.cn}]{\inits{J.}\fnm{Jiajia}~\lnm{Liu}\orcid{}}

\author[addressref={aff3}, email={fangsf@nssc.ac.cn}]{\inits{S.}\fnm{Shaofeng}~\lnm{Fang}\orcid{}}

\author[corref,addressref={aff1,aff4},email={rliu@ustc.edu.cn}]{\inits{R.}\fnm{Rui}~\lnm{Liu}\orcid{0000-0003-4618-4979}}

\address[id=aff1]{CAS Key Laboratory of Geospace Environment, Department of Geophysics and Planetary Sciences, University of Science and Technology of China, Hefei 230026, People’s Republic of China}
\address[id=aff2]{National Key Laboratory of Deep Space Exploration, School of Earth and Space Sciences, University of Science and Technology of China, Hefei 230026, China}
\address[id=aff3]{Space Science Big Data Technology Laboratory, National Space Science Center, Chinese Academy of Sciences, Beijing 100190, People’s Republic of China}
\address[id=aff4]{Mengcheng National Geophysical Observatory, University of Science and Technology of China, Hefei 230026, People’s Republic of China}

\runningauthor{J. Pan et al.}
\runningtitle{Solar Active Region Detection}

\begin{abstract}
Solar active regions are where sunspots are located and photospheric magnetic fluxes are concentrated, therefore being the sources of energetic eruptions in the solar atmosphere. The detection and statistics of solar active regions have been forefront topics in solar physics. In this study, we developed a solar active region detector (SARD) based on the advanced object detection model YOLOv8. First, we applied image processing techniques including thresholding and morphological operations to 6975 line-of-sight magnetograms from 2010 to 2019 at a cadence of 12~h, obtained by the Helioseismic and Magnetic Imager onboard the Solar Dynamic Observatory. With manual refinement, we labeled 26531 active regions in the dataset for further training and test with the detection model. Without any overlap between the training and test sets, the superior performance of SARD is demonstrated by an average precision rate as high as 94\%. We then performed a statistical analysis on the area and magnetic flux of the detected active regions, both of which yield log-normal distributions. This result sheds light on the underlying complexity and multi-scale nature of solar active regions.
\end{abstract}
\end{frontmatter}

\section{Introduction}
     \label{S-Introduction} 

A solar active region (AR) is known as an area where strong magnetic fields thread across the surface of the sun. Due to the substantial free energy contained in these magnetic fields, ARs serve as the primary source of diverse solar activities, such as flares and coronal mass ejections \cite[CMEs;][]{van2015evolution, liu2019evidence, liu2012slow}. These phenomena can subsequently lead to severe space weather events that negatively impact essential technological systems on Earth, e.g., threatening the safety of satellites and degrading the precision of global positioning systems \cite[]{dang2022unveiling}. 

In the past, long-term synoptic catalogs of ARs and sunspots were created through manual drawings, which has been replaced by computer-assisted analysis of solar images \cite[]{aschwanden2010image}. To date, there are two major data catalogs serving the solar and space physics community, namely the International Sunspot Numbers, which have been recorded daily since 1849, also known as the Wolf number or Zurich number \cite[]{10.1093/mnras/21.3.77}, and the National Oceanic and Atmospheric Administration (NOAA) AR catalog, which has been compiled by the US Air Force and the NOAA Space Weather Prediction Center. Each AR in the NOAA catalog is assigned a unique identification number, referred to as the NOAA AR number. The catalog provides information such as an AR’s location, longitudinal extent, sunspot area, and a three-letter classification for the sunspots, known as the  McIntosh classification \cite[]{mcintosh1990classification}. Previous studies on the statistical properties of sunspots and solar ARs include the work by \cite{bogdan1988distribution}, who identified a log-normal distribution for sunspot areas using data from the Mount Wilson Observatory, and the study by \cite{zharkov2005statistical}, who analyzed the size distribution of bipolar ARs on the Sun using daily full-disk magnetogram data from the National Solar Observatory/Kitt Peak.

In recent years, the successful launches of SOHO and SDO have enabled continuous, multi-channel, high-resolution observations of the Sun, significantly advancing our understanding of the solar dynamics. However, the large volume of solar images also necessitates the development of more objective and automated solar feature extraction methods to effectively process these images. There have already been several efforts to implement the automated identification of solar ARs. In the past, these mainly involve traditional image processing techniques, such as intensity threshold, morphological opening and closing operations, and region growing algorithms \cite[]{mcateer2005automated,colak2009automated, zhang2010statistical,caballero2014automatic,wang2023toward}. \cite{aschwanden2010image} reviewed four key image analysis techniques for the detection of solar event features, encompassing both spatial and temporal characteristics. These methods are: curvilinear one-dimensional (1D) feature detection, region-based two-dimensional (2D) feature detection, spectral methods, and artificial intelligence (AI)-based approaches. The curvilinear one-dimensional (1D) feature detection method is used to extract loop-like structures, such as coronal loops \cite[]{biskri2010extraction}. The region-based two-dimensional (2D) feature detection focuses on outlining solar events like sunspots and ARs, using traditional image processing techniques such as labeling, morphological operations, and watershed segmentation to extract features \cite[]{curto2008automatic}. Spectral methods decompose spatial images into Fourier components, identifying dominant spatial structures like ARs and filaments. It is effective in recognizing solar patterns through spectral analysis \cite[]{portier2001multiscale,ireland2008multiresolution}. The AI-based approaches,  particularly, neural networks, represent the most advanced method for solar feature detection. Trained on labeled data, neural networks are capable of automatically detecting various solar features including sunspots and ARs \cite[]{gong2024solar}.

With the advancement of object detection techniques in computer vision, numerous models have been proposed. From convolutional neural networks (CNNs) to Region-based CNN \cite[R-CNN;][]{girshick2014rich} and Faster R-CNN models \cite[]{ren2015faster}, these two-stage detectors have made significant contributions to the field of object detection. Normally, the first stage offers many region proposals, and the second stage performs object classification as well as bounding-box regression on each proposal to produce detections. These methods have also proven effective in solar feature detections; for example, \cite{baek2021solar} trained two representative object detection models, a single shot multi-box detector (SSD) and a Faster R-CNN model, both of which showed high performance in labeling coronal holes, sunspots, and prominences with bounding boxes. 

In contrast, one-stage object detection models, such as the YOLO series, view object detection task as a regression problem and predict object class and bounding box coordinates in a single process. \cite[]{redmon2016you,redmon2017yolo9000,bochkovskiy2020yolov4,li2022yolov6,wang2023yolov7}. As YOLO models evolve, they have demonstrated impressive performance in object detection, achieving higher accuracy than two-stage algorithms such as Faster R-CNN. 

In this study, we develop the SARD based on the YOLOv8 model, taking advantage of its outstanding performance in accuracy. The remaining content is organized as follows: Section \ref{S-general} details the data pre-processing steps, including AR labeling and necessary adjustments. The model architecture and detection results, including performance metrics, are presented in Sections \ref{S-features} and \ref{S-eandm}, respectively. Section \ref{S-statistics} focuses on the statistics of the detected ARs. Finally, Section \ref{S-discuss} provides concluding remarks.

\section{Data Preparation} 
      \label{S-general}      
\subsection{Pre-processing}
\label{preprocess}
The images utilized in this study are sourced from the Helioseismic and Magnetic Imager \cite[HMI;][]{Scherrer2012} onboard the Solar Dynamics Observatory \cite[SDO;][]{Pesnell2012} and can be accessed through the  Joint Science Operations Center (JSOC) repository\footnote{\url{http://jsoc.stanford.edu/}}. HMI has been operational since 1 May 2010. We employed the full-disk 720-s line-of-sight (LoS) magnetograms (hmi.M\_720s) to construct our datasets, covering the 24th solar cycle from 2010 to 2019. 

\begin{figure}    
\centerline{\includegraphics[width=1\textwidth,clip=]{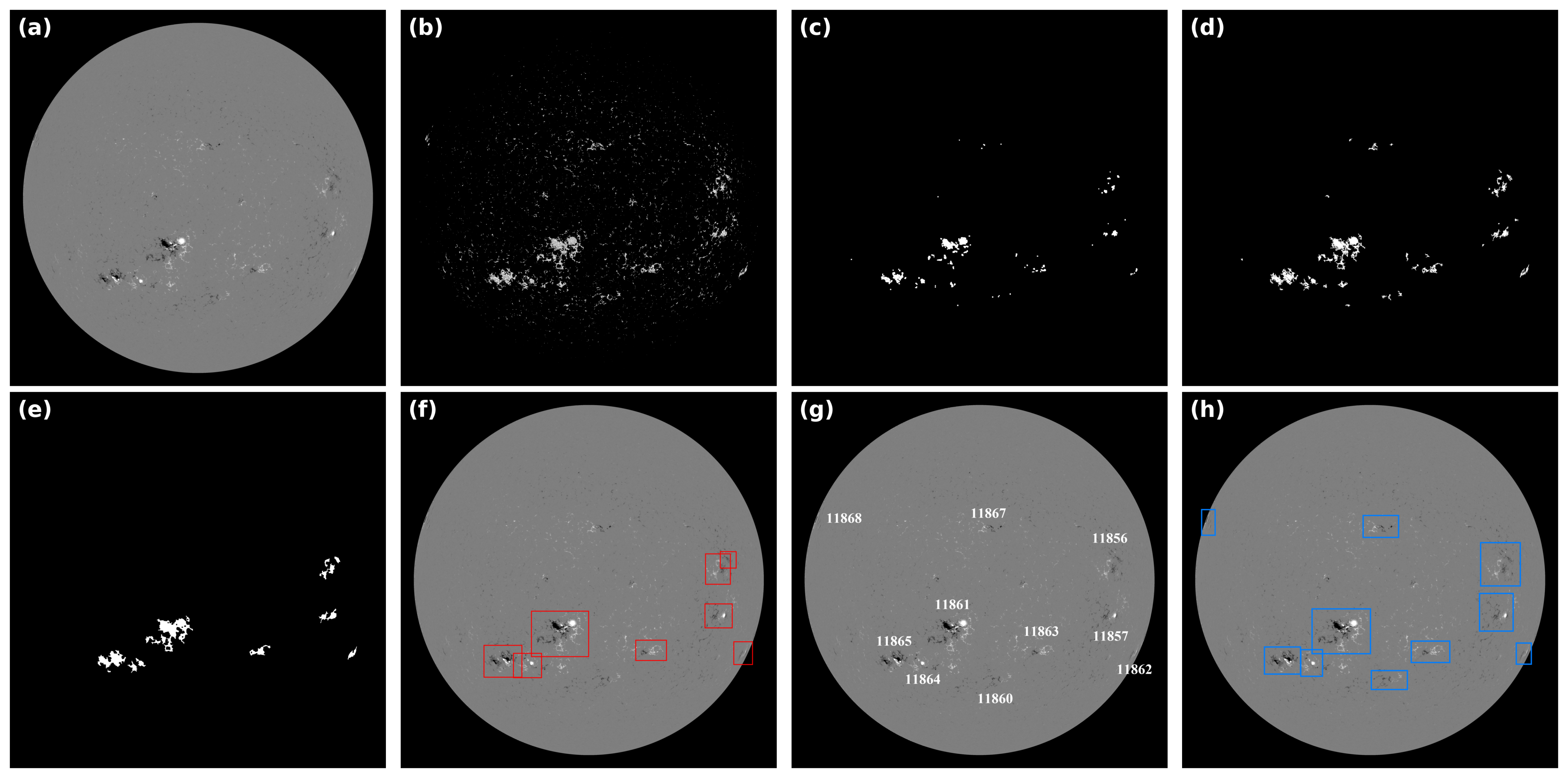}}
\small
        \caption{Flow chart of the image pre-processing during dataset preparation. Panel (a) shows the original LoS magnetogram of at 00:00:00 TAI on 12 October 2013. Panel (b) displays the result after OSTU thresholding. Panel (c) illustrates the outcome of the morphological opening operation, and Panel (d) shows the result of the region-growing method. Panel (e) presents the result of the morphological closing operation, while Panel (f) depicts the final detection of ARs with red bounding boxes. Panel (g) indicates the NOAA AR labels at 23:31:38 TAI on 11 October 2013, obtained from the Solar Monitor, and Panel (h) shows the AR labels after manual adjustment with blue bounding boxes.} 
\label{threshold}
\end{figure}

ARs must be labeled before being fed into the supervised deep learning model for training. To manual label such a large data set would be very time-consuming and tedious. Integrating advanced imaging processing and pattern recognition techniques \cite[]{wang2023toward, zhang2010statistical}, we carried out a four-step process to automate and streamline the labeling task (Figure~\ref{threshold}).

First (Figure~\ref{threshold}(b)), we applied threshold segmentation to isolate ARs from the surrounding background magnetic field based on intensity values. With the Otsu threshold method \cite[]{otsu1975threshold}, optimal threshold levels are automatically determined to ensure adaptive performance across varying image conditions. This segmentation method uses the inherent contrast differences in solar images, enabling the preliminary separation of regions of interest. 

Second (Figure~\ref{threshold}(c)), we employed morphological opening operations to refine the segmented regions; specifically, small noise particles and irrelevant features are removed with a $11\times11$ square structuring element. 

Third (Figure~\ref{threshold}(d)), the region-growing algorithm was implemented to recover the full size of an AR based on the pixels acquired from the previous step. This region-based technique begins with predefined seed points and iteratively expands them based on similarity criteria, such as intensity and texture, ensuring precise and contiguous region formation. Configured with an intensity threshold of 0.05, the region-growing algorithm recovered the pixels that are larger than threshold and connected to the seeds. 

Finally, a morphological closing operation was used to merge neighboring pixels and fill the gaps within the detected ARs (Figure~\ref{threshold}(e)). We then saturated the magnetogram to the range of $\pm650$~G, and set the vales of pixels beyond the solar limb to zeros. The final images labeled with ARs are shown in Figure~\ref{threshold}(f). \cite{wang2023toward} and \cite{zhang2010statistical} used the same procedure to identify ARs in synoptic magnetograms, achieving a recall rate of 73.8\%.  

It is not surprising that this automated detection procedure is sometimes at odds with the NOAA AR catalog. For example, the labels acquired from our pre-processing method for the solar disk on 12 October 2013 (Figure~\ref{threshold}(f)) are slightly different from those provided by the Solar Monitor\footnote{\url{www.SolarMonitor.org}} (Figure~\ref{threshold}(g)). Specifically, our procedure detects NOAA AR 11856 as two ARs, and fails to detect AR 11867 and AR 11860. We manually adjusted the labels to match the NOAA AR information (Figure~\ref{threshold}(h)) before feeding them to the model for training.

\subsection{Dataset Splitting}
We constructed the dataset using solar full-disk LoS magnetograms, with ARs being labeled through pre-processing (\S\ref{preprocess}). The dataset spans from 7 April 2010 to 31 December 2019. A total of 6,975 magnetograms were selected at a 12-hour cadence to ensure a comprehensive coverage of the 24th solar cycle. Within this dataset, 26531 ARs have been labeled, providing a robust basis for further analysis.

Given that the distribution of solar ARs remains highly similar over consecutive days, randomly splitting the dataset could result in data leakage. To mitigate this risk and ensure a comprehensive evaluation of the model’s performance, we designated three test sets, based on the number density and distribution of ARs (Table~\ref{split}). In addition, a 14-day gap---the transit time of active regions over the solar disk---is inserted in between the training and test datasets to ensure there are no overlaps between them.
    
\begin{table}[ht]
\centering
\caption{Dataset Splitting Strategy}
\label{split}
\begin{tabularx}{\textwidth}{>{\hsize=0.2\hsize\centering\arraybackslash}X >{\hsize=0.36\hsize\centering\arraybackslash}X >{\hsize=0.22\hsize\centering\arraybackslash}X >{\hsize=0.22\hsize\centering\arraybackslash}X}
\hline
\textbf{Dataset} & \textbf{Time Period} & \textbf{AR Number Density} & \textbf{Image Count} \\
\hline
\textbf{Training Set} & 
\begin{tabular}[c]{@{}c@{}}2011.7.1 to 2013.6.30 \\ 2014.1.1 to 2017.6.30 \\ 2018.7.1 to 2019.12.31\end{tabular} & 
--  & 5063\\
\hline
\textbf{Test 1 (T1)} & 2013.7.15 to 2013.12.17 & High & 311\\
\hline
\textbf{Test 2 (T2)} & 
\begin{tabular}[c]{@{}c@{}}Entire 2010 \\ 2011.1.1 to 2011.6.16\end{tabular} & 
Moderate & 790\\
\hline
\textbf{Test 3 (T3)} & 2017.7.15 to 2018.6.16 & Low & 699 \\
\hline
\end{tabularx}
\end{table}

\section{Method} 
      \label{S-features}      

\subsection{Basic Model Configuration} 

\begin{figure} [ht]
\centerline{\includegraphics[width=1\textwidth,clip=]{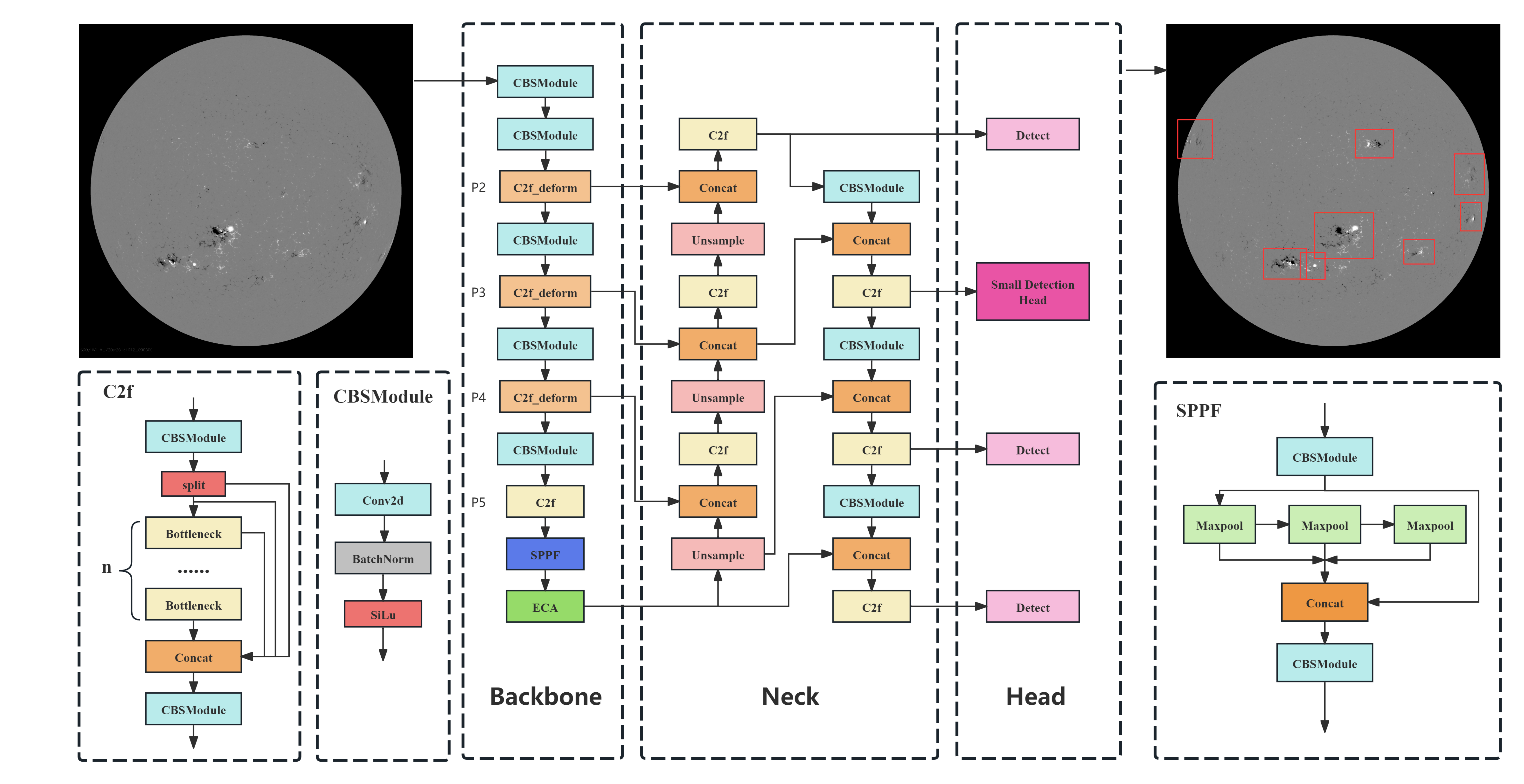}}
\small
        \caption{Configuration of SARD. The details of substructures are also illustrated.}
\label{yolov8}
\end{figure}

YOLOv8 is renowned for its balance between speed and accuracy. Building upon the strengths of its predecessors, YOLOv8 introduces several architectural enhancements that improve its performance on object detection tasks. The model mainly consists of three parts, i.e., backbone, neck, and head (Figure~\ref{yolov8}). The backbone is used to extract features from the input images by convolutional operations. The structure of backbone is similar to its predecessors, but modified with Cross Stage Partial (CSP) architecture to reduce computational complexity while maintaining accuracy \cite[]{wang2020cspnet}. Another enhancement is the C2f structure, which consists of two convolutional modules concatenated with several bottlenecks. It incorporates the benefits of ELAN structure \cite[]{wang2022designing} from YOLOv7 \cite[]{wang2023yolov7} to enhance detection accuracy. In addition, the basic convolution module is structured as a Convolution-Batch Normalization-SiLu module (CBSModule) to extract features from multi-scale feature maps. In the neck, YOLOv8 employs a Path Aggregation Network \cite[PAN;][]{liu2018path} and a Feature Pyramid Network \cite[FPN;][]{lin2017feature} to better aggregate multi-scale features, improving the detection of objects of various sizes. The model also utilizes an anchor-free detection head and decouples the classification and detection process, enabling independent optimization of object detection and classification, which leads to improvements in the overall detection accuracy \cite[]{duan2019centernet}. These enhancements make YOLOv8 an ideal framework for our task, offering the necessary precision and computational efficiency for detecting ARs in solar full-disk magnetograms.

\subsection{Deformable Convolution}
\label{DCN}
The C2f module is a fundamental building block in YOLOv8, which is designed to enhance feature extraction and representation. It consists of two convolutional modules and multiple bottlenecks. However, the common convolution kernels in this module restrict the receptive field of the network, and only capture local feature information \cite[]{WEI2022106653}. Deformable convolutional networks (DCN) introduce learnable offsets, enabling convolutional neural networks to adaptively adjust their receptive fields based on the input images \cite[]{dai2017deformable}. This adaptability enhances the ability of convolution modules to recognize complex spatial relationships and objects with irregular edges. Passing the input feature map $\chi$ through a traditional convolutional layer gives the output at the position \( p \) as follows,
\begin{equation}
    y(p) = \sum_{k} w_k \cdot \chi(p + p_k),
\end{equation}
where \( k \) is the number of convolutional kernels, \( w_k \) are the learnable weights, and \( p_k \) are the fixed relative positions in the grid. 

Deformable convolution modifies this operation by introducing offsets:
\begin{equation}
    y(p) = \sum_{k} w_k \cdot \chi(p + p_k + \Delta p_k),
\end{equation}
where \( p_k \) represents the relative positions of the convolutional kernel, and \( \Delta p_k \) is the offsets that adjust these positions based on the input feature map. By integrating DCN with the C2f module, the limitations of convolutional kernels, which are confined to fixed and regular grid sampling, are effectively addressed. The structure of this integrated module is shown in Figure~\ref{yolov8}. The incorporation of DCN enables the C2f module to model complex spatial relationships more accurately and adapt to objects with irregular edges. This enhanced flexibility is particularly suitable for detecting solar ARs, which usually hold irregular boundaries and various shapes.

\subsection{Attention Module}
\label{subsec:eca}
ARs that are densely distributed pose a challenge to the detection model. In this regard, we incorporated a lightweight and efficient attention mechanism, namely Efficient Channel Attention (ECA), into the backbone of YOLOv8 to enhance the model's ability of detecting small-scale and closely clustered ARs. The attention module selectively emphasizes such features by assigning more weights to them, while suppressing the less important background. This improves the detection accuracy and therefore the overall performance of the model.

\begin{figure} [ht]
\centerline{\includegraphics[width=1\textwidth,clip=]{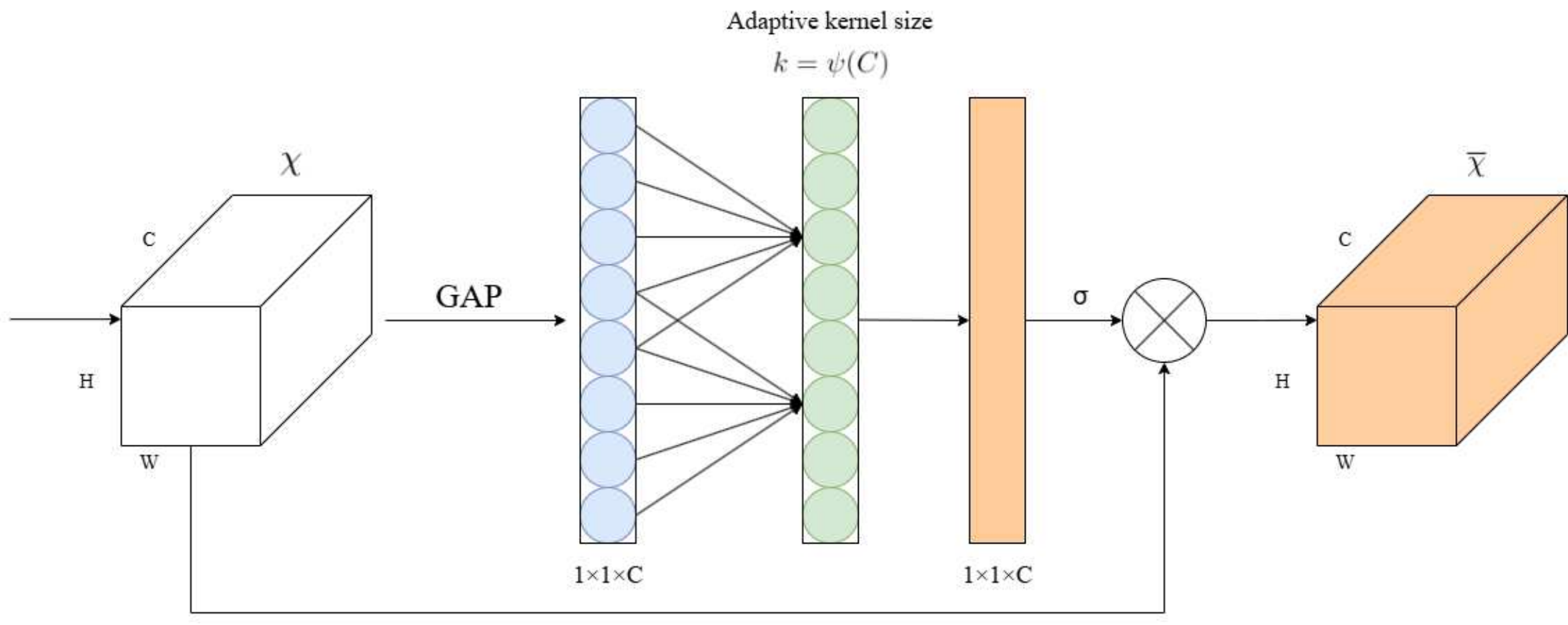}}
\small
        \caption{Configuration of Efficient Channel Attention (ECA) mechanism. }
\label{fig:eca}
\end{figure}

Unlike traditional channel attention mechanisms, ECA avoids the complexity of dimensionality reduction and captures local cross-channel interactions using a fast 1D convolution \cite[Figure~\ref{fig:eca};][]{wang2020eca}. If there is an output feature map $\chi \in \mathbb{R}^{W \times H \times C}$ from a convolutional block, ECA first applies a global average pooling (GAP):
\begin{equation}
    g(\chi) = \frac{1}{W\times H} \sum_{i=1}^{W} \sum_{j=1}^{H} \chi_{ij}.
\end{equation}
These features are then passed through a 1D convolution $\text{Conv}_k(g(\chi))$ with a kernel size $k$. In order to adaptively determine $k$, the author introduced a mapping $\phi$ between channel dimension $C$ and kernel size $k$, i.e., $C=\phi(k)=2^{\gamma \times k-b}$, inspired by the philosophy of group convolutions \cite[]{xie2017aggregated}. Then, $k$ is adaptively determined by a function $\psi$ dependent on the channel dimension $C$ to effectively capture local cross-channel interactions \cite[]{wang2020eca}, i.e. ,
\begin{equation}
    k = \psi(C) = \left\vert\frac{\log_{2}C}{\gamma}+\frac{b}{\gamma}\right\vert_\mathrm{odd},
\end{equation}
where the subscript `odd' indicates an operation of taking the nearest number to maintain the symmetry of the convolution kernel. The values of $\gamma$ and $b$ are set to 2 and 1, following \cite{wang2020eca}. Then a sigmoid activation function $\sigma(x)=1/(1+e^{-x})$ is used to generate channel-wise weights $\omega$: 
\begin{equation}
    \omega = \sigma(\text{Conv}_k(g(\chi))).
\end{equation}
Finally, the re-calibrated output feature map is obtained by an element-wise multiplication (the operator $\otimes$ in Figure \ref{fig:eca}) between the weight and the input feature map $\chi$, i.e.,
\begin{equation}
   \bar{\chi} = \omega \otimes \chi.
\end{equation}

\subsection{Detection Head} 
\label{SS-head}
In the original YOLOv8 model, with an input image size of 1024$\times$1024, the detection process depends on three different scales of feature maps. Specifically, object detection occurs at the resolution of 128$\times$128 for small objects, 64$\times$64 for medium-scale objects, and 32$\times$32 for large objects. In order to reduce missed detections and improve the performance of the model on small-scale objects, an additional detection head is added to the detection module sized at 256$\times$256. Consequently, the improved model works on four different scales in the detection module (Figure~\ref{yolov8}). 

\section{Experiment and Metrics}
\label{S-eandm}
\subsection{Evaluation Metrics}
SARD is designed for automatically extracting ARs from solar full-disk LoS magnetograms. The output of the model is the boxes bounding ARs. A detection is determined to be true or false depending on the intersection over union (IoU) between the predicted bounding box and the ground truth, the latter of which is the actual location of the AR as annotated in magnetograms (\S\ref{S-general}). Specifically, a detection is considered correct if the IoU exceeds 50\%. In order to validate the model, the metrics including precision, recall, F1-score, and average precision (AP) are used. The precision (P) and recall (R) are defined as follows:
\begin{equation}
    P=\frac{TP}{TP+FP};
\end{equation}
\begin{equation}
    R=\frac{TP}{TP+FN},
\end{equation}
where TP denotes true positives, representing the samples that are correctly detected as positive samples; FP represents false positives, which are negative samples incorrectly identified as positive; and FN represents false negatives, which are positive samples incorrectly identified as negative. 

\begin{figure}
\centerline{\includegraphics[width=1\textwidth,clip=]{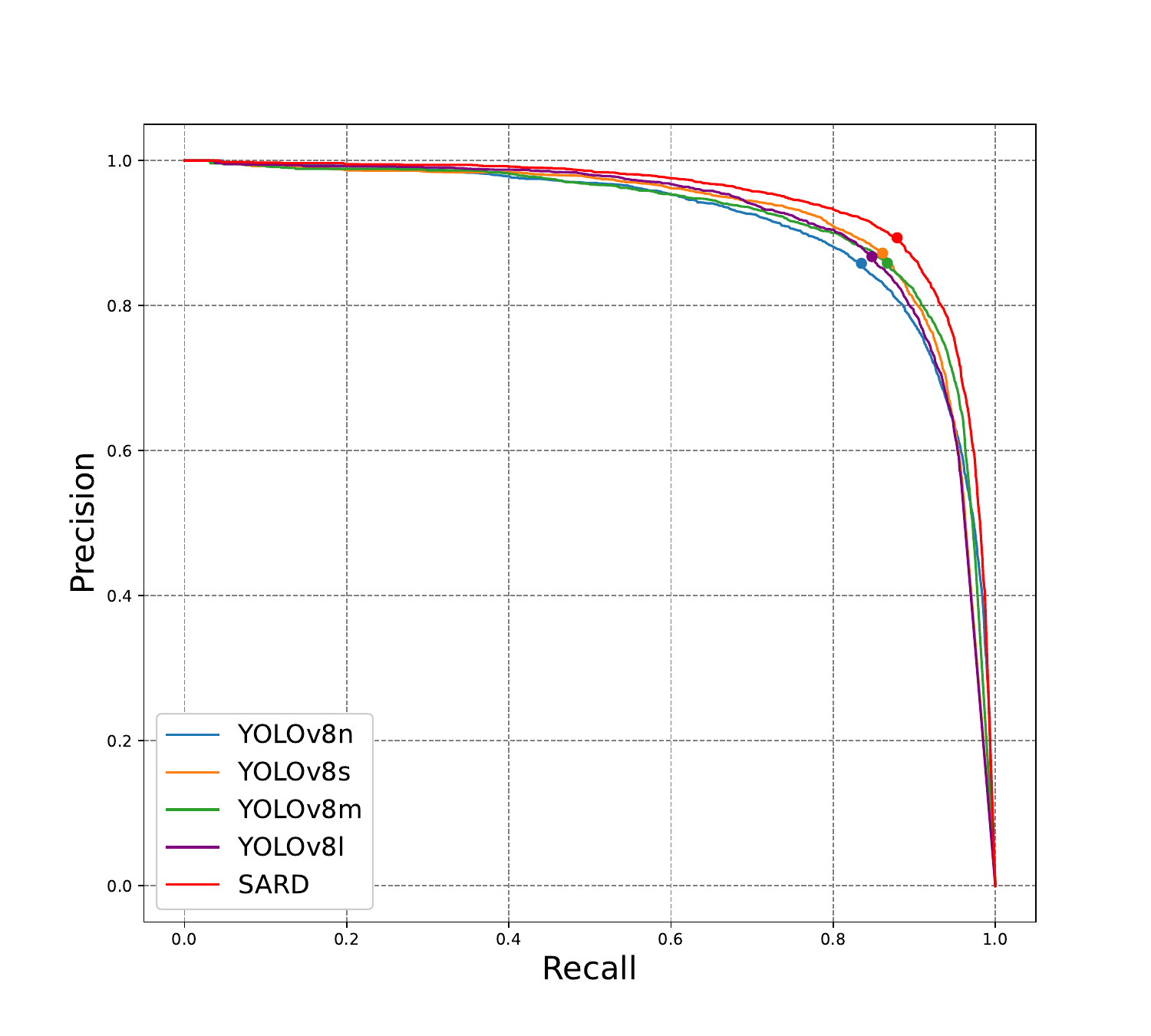}}
\small
        \caption{The P-R curves of YOLOv8n, YOLOv8s, YOLOv8m, YOLOv8l, and SARD on the entire test set. P and R are calculated at different confidence thresholds, and the area under the curve is the AP. The dots indicate the final reported values of precision and recall.}
\label{prcurve}
\end{figure}

With precision and recall, we can calculate the $F1$-score, a harmonic mean of the two:
\begin{equation}
    F1=\frac{2P\times R}{P+R}.
\end{equation}
To obtain a comprehensive evaluation, we rank all the detections based on their confidence scores. The confidence score is a value between 0 and 1 that represents the model's certainty about each detection, quantifying the likelihood that a predicted bounding box accurately corresponds to an object instance. By applying a threshold to the confidence scores, we compute a precision-recall (P–R) pair, and by varying the threshold, we generate the P–R curve. The final reported precision and recall values (Table~\ref{tbl:metrics}) are those obtained at the confidence threshold that maximizes the $F1$-score. The AP is then derived by integrating along the P-R curve, which reflects the overall robustness and performance of the model. Figure \ref{prcurve} shows the P-R curves of YOLOv8n, YOLOv8s, YOLOv8m, YOLOv8l, and SARD on the entire test set, and the dots indicate the final reported values of precision and recall.

\subsection{Experimental results}
\label{SS-exresults}

YOLOv8 provides a tiered model family YOLOv8n, YOLOv8s, YOLOv8m and YOLOv8l by scaling two architectural hyper-parameters: width (channels per layer) and depth (number of layers). Moving from n to l increases these values, yielding more parameters and computational resource consumption. However, in our experiments, we observed that YOLOv8s produced the best results on our data set, suggesting that it strikes an optimal balance between model capacity and generalization. During the training stage, some of the key hyper-parameters are shown in Table~\ref{tbl:hyperparameters}, and we kept 20\% 
 of the training set as a validation set for hyper-parameter tuning and model structure selection. Additionally, data augmentation techniques, including mosaic and image scaling, were applied to enhance the robustness of the model.

\begin{table}[ht]
\caption{Hyperparameters setup during training stage.}
\label{tbl:hyperparameters}
\begin{tabularx}{\textwidth}{>{\centering\arraybackslash}X>{\centering\arraybackslash}X}
\hline
Hyperparameters & Setup \\
\hline
Epochs & 100 \\
Momentum & 0.937 \\
Warm up epochs & 3 \\
Initial learning rate & 0.001 \\
Weight decay & 0.0005 \\
Batch size & 16 \\
Input image size & 1024 $\times$ 1024 \\
Optimizer & AdamW \\
\hline
\end{tabularx}
\begin{tablenotes}[flushleft]
\footnotesize
\item \textit{Note.} see \cite{kingma2014adam} for AdamW optimizer.
\end{tablenotes}
\end{table}

The performance metrics of all the models mentioned above, including P, R, AP, and F1-score, are presented in Table \ref{tbl:metrics}, in which the performance of the traditional image processing method in \cite{zhang2010statistical} is also cited as a reference. Compared with \cite{zhang2010statistical}, which achieved a precision of 0.778 and a recall of 0.738 as an average performance mentioned in their paper, SARD not only achieved a better performance but also shortened the detection time from tens of seconds per magnetogram to under one second. The YOLOv8s and YOLOv8m model performed the best on the test data set in terms of AP and F1-score. Despite their greater capacity, larger models like YOLOv8l may not be suited well to our specific data set. The larger the parameters (width and depth) of the model, the greater the computational resources are required, and the longer the inference time would be consumed. In terms of precision, recall, AP, and F1-score, SARD based on YOLOv8s is superior to the original ones (Table~\ref{tbl:metrics}), as well as the latest versions of the YOLO series, such as YOLOv11 and YOLOv12, because certain improvements in these versions are not well-suited to the characteristics of our task. Moreover, \cite{quan2021solar} applied the YOLOv3 model to detect solar ARs in LoS magnetograms, achieving a recall of 94\%. However, their test set consisted of images from the solar minimum period, when there are few ARs, potentially making the detection task easier.

\begin{figure}
\centerline{\includegraphics[width=1\textwidth,clip=]{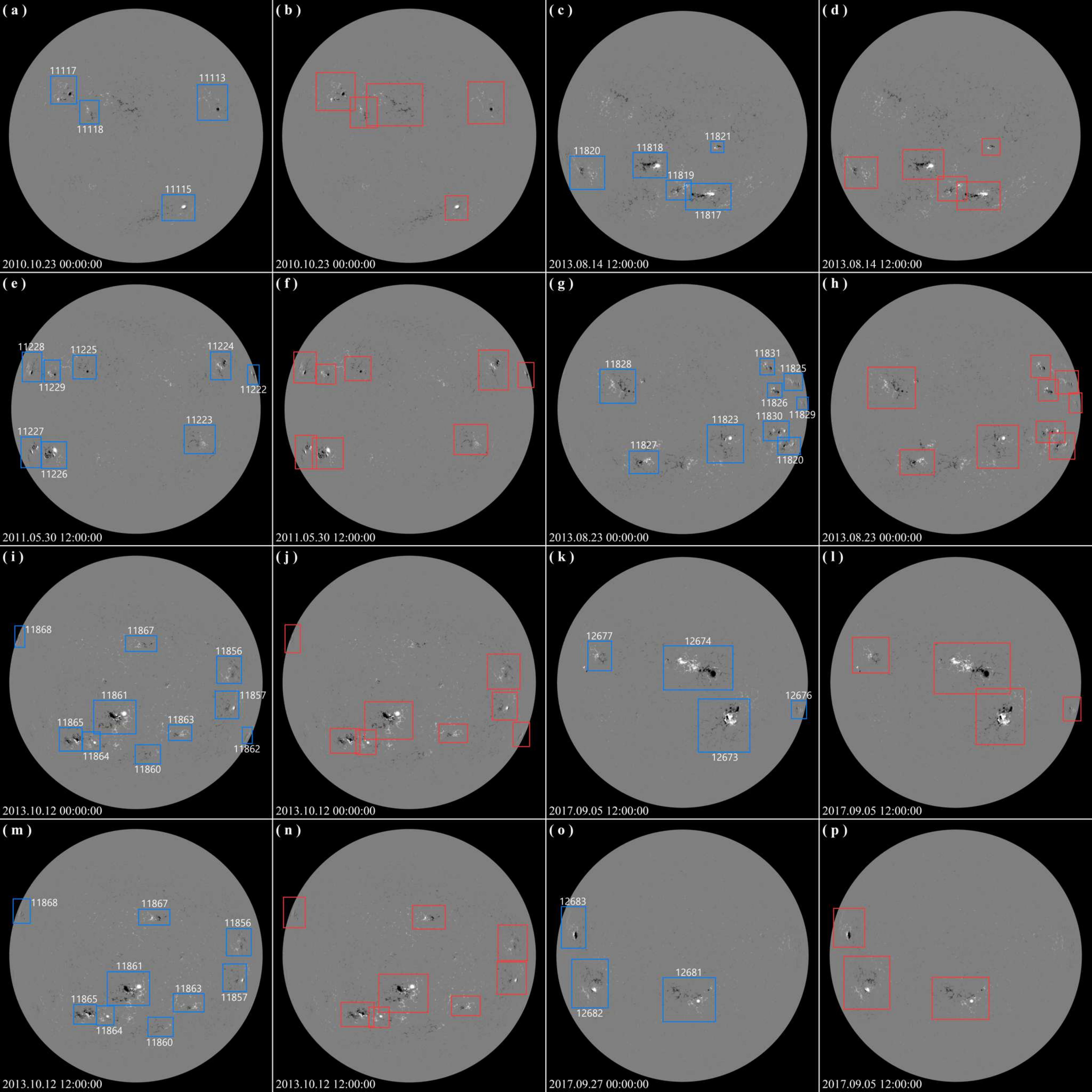}}
\small
        \caption{Examples of ARs detected by the SARD using LoS magnetograms. NOAA ARs are artificially marked by blue boxes with the AR numbers annotated. The red bounding boxes are given by the SARD.}
\label{deresult}
\end{figure}

Figure~\ref{deresult} shows eight examples from the test data set. The red bounding boxes are given by SARD, and blue ones marked NOAA ARs based on the information given by the Solar Monitor. Overall, SARD successfully detects most of the ARs, including the clustered regions, ARs 11856, 11861, 11864, and 11865. For the magnetogram on 12 October 2010, SARD correctly separates the AR 11856 from its neighbors, while the traditional image processing pipeline mistakenly merges them into a single AR (see also \S\ref{preprocess}). However, two decayed ARs, AR 11860 and AR 11867, are not detected. For the magnetogram on 23 October 2010, when ARs are distributed sparsely, the model successfully detected all the ARs, except that a diffuse region with strong fields near the disk center is mistakenly detected as an AR.

\begin{table}[ht]
\caption{Comparison of model performance among YOLOv8n, YOLOv8s, YOLOv8m, YOLOv8l, YOLOv11s, YOLOv12s, SARD, and the results from \cite{zhang2010statistical}.}
\label{tbl:metrics}
\begin{tabularx}{\textwidth}{>{\centering\arraybackslash}X>{\centering\arraybackslash}X>{\centering\arraybackslash}X>{\centering\arraybackslash}X>{\centering\arraybackslash}X}
\hline
Model & Precision & Recall & F1-score & AP\\
\hline
\cite{zhang2010statistical} & 0.778 & 0.738 & 0.758 & -- \\
YOLOv8n & 0.845 & 0.844 & 0.845 & 0.916 \\
YOLOv8s & 0.874 & 0.859 & 0.866 & 0.921 \\
YOLOv8m & 0.863  & 0.861 & 0.862 & 0.921 \\
YOLOv8l & 0.874  & 0.841 & 0.857 & 0.919 \\
YOLOv11s & 0.852 & 0.833 & 0.842 & 0.921 \\
YOLOv12s & 0.853 & 0.849 & 0.851 & 0.92 \\
\textbf{SARD} & \textbf{0.89} & \textbf{0.881} & \textbf{0.886} & \textbf{0.943} \\
\hline
\end{tabularx}
\end{table}

Table \ref{devide} lists the performance of the model on the test datasets T1, T2 and T3, respectively. T1 consists of magnetograms from the year of 2013, the peak of the 24th solar cycle. During this period, ARs of various sizes are densely distributed on the solar surface. SARD achieved an impressive performance on T1, with a precision of 0.896, a recall of 0.901, an AP of 0.95, and an F1-score of 0.898. These results demonstrate the model's capability in handling the complex distribution of ARs. SARD's performance was slightly degraded with T2 and T3, in which ARs are sparsely distributed. In particular, T3 corresponds to a period of low solar activity, reflecting SARD's tendency to misclassify quiet regions with relatively strong flux as ARs during periods of low solar activity. For example, the diffuse region near the disk center on 10 October 23 is falsely recognized as an AR (Figure \ref{deresult}). Nevertheless, the model's performance across the entire dataset is excellent, with a precision of 0.895, a recall of 0.884, an AP of 0.946, and an F1-score of 0.889. These results demonstrate the model's robustness in detecting solar ARs of varying sizes and densities over the entire cycle.

\begin{table}[ht]
\caption{Model performance on the test sets.}
\label{devide}
\begin{tabularx}{\textwidth}{>{\centering\arraybackslash}X>{\centering\arraybackslash}X>{\centering\arraybackslash}X>{\centering\arraybackslash}X>{\centering\arraybackslash}X>{\centering\arraybackslash}X}
\hline
Test set & Precision & Recall & F1-score & AP & AR Number Density \\
\hline
Test 1 (T1) & 0.893 & 0.898 & 0.896 & 0.947 & High \\
Test 2 (T2) & 0.888 & 0.874 & 0.881 & 0.944 & Moderate \\
Test 3 (T3) & 0.864  & 0.891 & 0.877 & 0.94 & Low \\
Overall           & 0.89 & 0.881 & 0.886 & 0.943 & -- \\
\hline
\end{tabularx}
\end{table}

To further validate the effectiveness of the proposed improvements to the base YOLOv8 model, an ablation experiment was conducted by incrementally integrating the individual enhancement module over the base model. The results of this experiment are presented in Table \ref{tbl:abl}. The term "YOLOv8s+head" represents the base model augmented with a detection head designed to improve performance on small-scale objects. The addition of the ECA module enhances the model’s ability to focus on ARs by applying more emphasis on the small-scale features; the further integration of the DCN module is particularly effective for detecting objects with irregular shapes, making it well-suited for ARs. Compared to the baseline model, the addition of the detection head alone yields a 0.2\% increase in AP. The incorporating the ECA module and DCN results in additional improvements of 1.1\% and 0.9\% in AP, respectively, alongside enhanced recall, precision, and F1-score.

\begin{table}[ht]
\caption{Ablation experiments to show the incremental improvements over YOLOv8 leading to the SARD}
\label{tbl:abl}
\begin{tabularx}{\textwidth}{>{\hsize=.52\hsize\RaggedRight\arraybackslash}X>{\hsize=.12\hsize\centering\arraybackslash}X>{\hsize=.12\hsize\centering\arraybackslash}X >{\hsize=.12\hsize\centering\arraybackslash}X>{\hsize=.12\hsize\centering\arraybackslash}X}

\hline
Model & Precision & Recall & F1-score & AP \\
\hline
YOLOv8s & 0.875 & 0.859 & 0.867 & 0.921 \\
YOLOv8s+head & 0.858 & 0.855 & 0.857 & 0.923 \\
YOLOv8s+head+ECA & 0.868 & 0.868 & 0.868 & 0.934 \\
YOLOv8s+head+ECA+DCN (SARD)  & 0.89 & 0.881 & 0.886 & 0.943 \\
\hline
\end{tabularx}
\end{table}

\section{Statistics} 
\label{S-statistics}

In this section, we analyze the statistical properties of solar ARs over the period from 2010 to 2019, using the dataset consisting of 6,975 solar full-disk LoS magnetograms, at a time cadence of 12 hours. To minimize the projection effects and avoid duplicate counting, we selected a total of 1,331 ARs by applying a narrow longitudinal window of $\pm3.5\degree$ from the central meridian to the magnetograms. Only those ARs whose center falls within this window are selected for further analysis. Although this criterion would exclude a few dozen newly emerging ARs, the statistics is barely affected.  To obtain the magnetic flux of ARs, we took the LoS component of the phtospheric magnetic field as a good approximation of the vertical component, and performed a cylindrical equal-area (CEA) projection \cite[]{thompson2006coordinate,caballero2014automatic} to obtain the area of the quadrilateral bounding each individual AR.

\begin{figure} [ht]
\centerline{\includegraphics[width=1\textwidth,clip=]{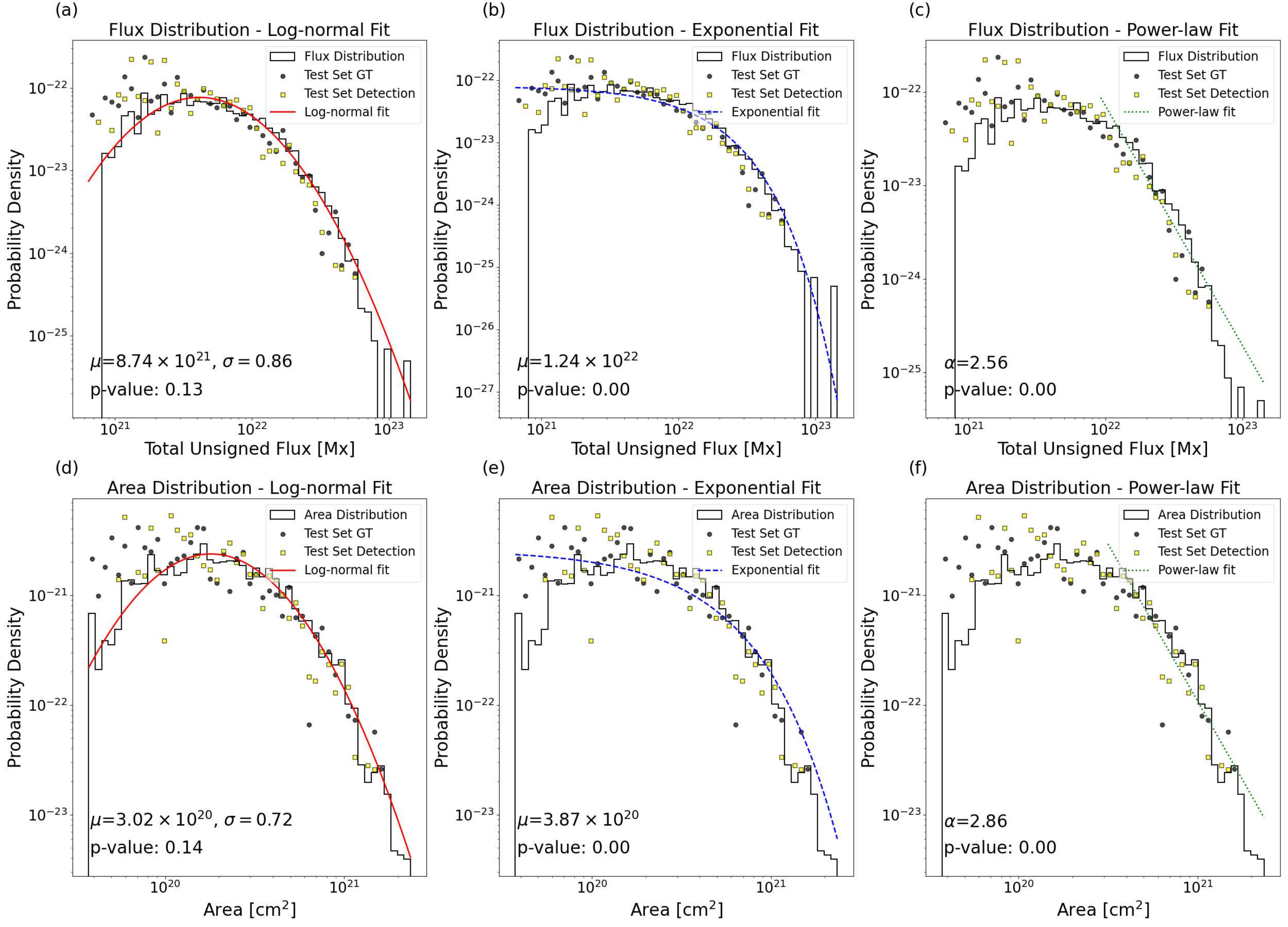}}
\small
        \caption{The flux and area distributions in log-log scales. The histograms represent the distributions of the entire dataset, while the scatter points represent the test set, with black and yellow colors indicating the ground truth (GT) and the SARD-detected results, respectively. The fitting parameters for the entire dataset are displayed in the bottom left of each panel.} Panels (a)-(c): distribution and fitting results of flux of the ARs. Panels (d)-(f): distribution and fitting results of the area of ARs. The red, blue, and green lines represent the log-normal, exponential, and power-law fit, respectively.
\label{fa_distribution} 
\end{figure}

Figure \ref{fa_distribution} presents the flux (a--c) and area (d--f) distributions of all selected ARs, displayed in histograms in a log-log scale. As a reference, the ground truth and SARD-detected results from the test set are also shown as black and yellow points, respectively. One can see that these distributions deviate from normal distributions. We fit the distributions with a log-normal (red), an exponential (blue), and a power-law function (green), respectively; i.e., 
\begin{equation}
\begin{split}
     f(x;\mu, \sigma)=\frac{1}{x\sigma\sqrt{2\pi}}\exp\left(-\frac{\ln({x}/{\mu})^{2}}{2\sigma^{2}}\right),\\
     f(x;\mu)=\exp\left(-\frac{x}{\mu}\right), \\
     f(x;\alpha)=\frac{\alpha-1}{x_\mathrm{min}}\left(\frac{x}{x_\mathrm{min}}\right)^{-\alpha},
\end{split}
\end{equation}
where $\mu$ represents the expected value, $\sigma$ is the standard deviation, and $\alpha$ is the exponent in the power law. The power-law fit was only performed on the tail of the distributions, which appears to follow power laws, so the value of $x_\mathrm{min}$ is fixed to the median value of the data. The fitting and Kolmogorov-Smirnov (KS) test results of the entire data set are also provided in Figure~\ref{fa_distribution}. 

The results of the KS test further support the conclusion that the log-normal function provides the best fit for both the area and flux distributions for the entire data set, among the three functions. Specifically, for the flux distribution, the $p$-value for the log-normal distribution is 0.13, which exceeds the conventional significance level of 0.05, indicating no significant deviation from the log-normal fit. In contrast, the KS test for both the exponential distribution and power-law yields a $p$-value close to 0, suggesting a poor fit to the data. Similar results are obtained for the area distribution. 

As an experiment for comparison, we selected 310 SARD-detected and 307 ground-truth ARs, respectively, from the test set under the same criterion. Not only are their flux and area distributions very similar to each other, but also similar to those from the entire dataset. For example, the KS test of the log-normal fitting for the SARD-detected ARs returns a $p$-value of 0.59 and 0.38 for the flux and area distribution, respectively, while those for the expoential and power-law functions are well below 0.05, indicating that the log-normal function again provides the best fit. This experiment demonstrates that the flux and area distributions of ARs are robustly log-normal, and that the SARD is reliable.

The normal distribution of AR areas were reported by \cite{bogdan1988distribution} and \cite{ograpishvili1994distribution}. However, several studies have observed that the magnetic flux distribution of ARs follows an exponential law, as suggested by \cite{tang1984statistical} and \cite{schrijver1997modeling}. Furthermore, \cite{das1982distribution} found that the distribution of sunspot areas follows a log-normal distribution, while the flux of sunspots follows a power law. \cite{ograpishvili1994distribution}, using data from the Abastumani Astrophysical Observatory Solar Service, reported that both the area and importance (defined as the product of area and brightness intensity) distributions of ARs conform to a log-normal distribution. Similarly, \cite{bogdan1988distribution} utilized data from the Mount Wilson white-light plate collection and found that the distribution of sunspot umbral areas follows a log-normal distribution. \cite{bogdan1988distribution} proposed that the observed log-normal size distribution might be linked to the fragmentation of magnetic flux elements in the solar envelope, where the magnetic elements undergo random subdivisions, leading to a size distribution that follows a log-normal pattern. Our findings are in agreement with those of \cite{bogdan1988distribution}, supporting the idea that the log-normal distribution is indicative of fragmentation processes, which are inherent to the global dynamo that is interacting with turbulent flows in the convection zone.

\section{Discussion and Conclusion}
\label{S-discuss}
In this study, we proposed an efficient and accurate object detection model, SARD, for the solar AR detection with full-disk LoS magnetograms. To save the time and labor required for labeling ARs for the training set, we automated the process by using advanced image processing techniques such as intensity thresholding, morphological operations, and region-growing algorithms. After manual adjustment with reference to the NOAA AR catalog, the dataset, which includes a total of 6,975 full-disk magnetograms and 19,098 labeled ARs from 2010 to 2019, provides a solid basis for model training and test.

The SARD is based on the YOLOv8 model because of its proven ability to deliver high performance in object detection tasks, with a good balance between speed and accuracy. By incorporating DCN and the ECA module, we improved the model’s ability to handle irregular and multi-scaled solar ARs. The results showed that the SARD achieved a remarkable performance, with the precision of 0.895, the recall of 0.884, and the AP of 0.946, demonstrating its efficacy in detecting solar ARs with high accuracy. Compared with traditional image-processing pipelines, SARD offers both higher detection accuracy and faster detection speed, making it better suited for large-scale data processing (\S\ref{SS-exresults}).

The statistical analysis suggests that both the area and magnetic flux distribution of the detected ARs is best fitted by a log-normal distribution, which is consistent with previous studies, such as \cite{bogdan1988distribution} and \cite{ograpishvili1994distribution}. The log-normal distribution could be explained by the presence of a fragmentation process in the Sun's magnetic flux elements.

To conclude, we constructed a robust model for solar AR detection and offered new insights into the statistical properties of ARs. The use of deep learning models, particularly YOLOv8, has proven effective in automating the detection process, making it conducive to analyzing large amounts of solar data that are challenging for manual handling. Furthermore, the validation of the log-normal distribution of AR area and flux is consistent with the global dynamo model, in which the solar magnetic activity is driven by fragmentation processes occurring within the convection zone.

\begin{acknowledgments}
    The authors are thankful to SolarMonitor for providing the NOAA AR annotation data and to the Joint Science Operations Center (JSOC) at Stanford University for providing the solar full-disk LoS magnetograms used in this study.
\end{acknowledgments}

\begin{authorcontribution}
    R.L. devised the research plan, supervised the study, revised and finalized the manuscript. J.P. carried out the study under the guidance of R.L., J.L., and S.F.,  and wrote the draft. All authors reviewed the manuscript.
\end{authorcontribution}

\begin{fundinginformation}
This work was supported by the National Key $R\&D$ Program of China (2022YFF0711402) and the National Natural Science Foundation of China (NSFC; 12373064, 42188101, 11925302).
\end{fundinginformation}

\begin{dataavailability}
    The dataset generated and analyzed in this study are available at \url{https://rec.ustc.edu.cn/share/c0a85070-1098-11f0-9e80-07981db8dd4b}
\end{dataavailability}

\begin{materialsavailability}
    The codes and the trained models are available at \url{https://github.com/WhCanI/AR_DETECT.git}
\end{materialsavailability}


\bibliographystyle{spr-mp-sola}
\bibliography{sola_example_6}  

\IfFileExists{\jobname.bbl}{} {\typeout{}
\typeout{****************************************************}
\typeout{****************************************************}
\typeout{** Please run "bibtex \jobname" to obtain} \typeout{**
the bibliography and then re-run LaTeX} \typeout{** twice to fix
the references !}
\typeout{****************************************************}
\typeout{****************************************************}
\typeout{}}

\end{document}